\documentclass[prl,aps,reprint,amsmath,amssymb,nofootinbib,superscriptaddress,10pt]{revtex4-1}

\usepackage{graphicx,dcolumn,bm}
\usepackage{amsmath}
\usepackage{amsthm}
\usepackage{amsfonts}
\usepackage{mathrsfs}
\usepackage[usenames,dvipsnames]{xcolor}
\usepackage[colorlinks=true,citecolor=MidnightBlue,linkcolor=MidnightBlue,urlcolor=MidnightBlue]{hyperref}
\usepackage{braket}
\usepackage{bbm}

\renewcommand{\vec}[1]{{\bf #1}}

\begin{document}
\title{Directional Dicke Subradiance with Nonclassical and Classical Light Sources}

\author{Daniel~Bhatti}
\author{Raimund~Schneider}
\affiliation{Institut f\"{u}r Optik, Information und Photonik, Universit\"{a}t Erlangen-N\"{u}rnberg, 91058 Erlangen, Germany}
\affiliation{Erlangen Graduate School in Advanced Optical Technologies (SAOT), Universit\"at Erlangen-N\"urnberg, 91052 Erlangen, Germany}
\author{Steffen~Oppel}
\affiliation{Institut f\"{u}r Optik, Information und Photonik, Universit\"{a}t Erlangen-N\"{u}rnberg, 91058 Erlangen, Germany}
\author{Joachim~von~Zanthier}
\affiliation{Institut f\"{u}r Optik, Information und Photonik, Universit\"{a}t Erlangen-N\"{u}rnberg, 91058 Erlangen, Germany}
\affiliation{Erlangen Graduate School in Advanced Optical Technologies (SAOT), Universit\"at Erlangen-N\"urnberg, 91052 Erlangen, Germany}


\date{\today}

\begin{abstract}


We investigate Dicke subradiance of distant quantum sources in free space, i.e., the spatial emission pattern of spontaneously radiating non-interacting multi-level atoms or multi-photon sources, prepared in totally antisymmetric states. We find that the radiated intensity is marked by a strong suppression of spontaneous emission in particular directions.
In resemblance  to the analogous, yet inverted, superradiant emission profiles of $N$ distant two-level atoms prepared in symmetric Dicke states, we call the corresponding emission pattern  \textit{directional Dicke subradiance}.
We also show that higher order intensity correlations of the light incoherently emitted by 
statistically independent thermal light sources display 
the same directional Dicke subradiant behavior. We present measurements of directional Dicke subradiance for $N = 2, \ldots, 5$ distant thermal light sources corroborating the theoretical findings.

\end{abstract}

%

\maketitle


Dicke superradiance, i.e., the enhanced spontaneous emission in space and time of atoms in highly entangled symmetric Dicke states, has been extensively studied over the last 60 years
\cite{Dicke(1954),Eberly(1971),Manassah(1973),AGARWAL(1974),Haroche(1982),Scully(2006),Scully(2007),
Scully(2009)Super,Kaiser(2012),Kaiser(2013),
Cirac(2011),
Rohlsberger(2010),
Wiegner(2011)PRA,Oppel(2014),Wiegner(2015),Bhatti(2016)}.
In contrast, its cryptic twin, subradiance, has been much less investigated, mainly due to its higher degree of complexity and increased demands for experimental verification, even though considerable progress has been made recently. Since the first indirect observation 
\cite{Pavolini(1985)}, the main focus 
has been on studying subradiance of two two-level atoms \cite{DeVoe(1996), Hettich(2002), Barnes(2005), Julienne(2012),McGuyer(2015)}. This configuration 
is most transparent, less fragile \cite{Woggon(2005)}, and, moreover, can be prepared in both parities, a fully symmetric as well as a fully antisymmetric state, where the latter decouples entirely from the vacuum field for small atom separations \cite{AGARWAL(1974),Haroche(1982)}. For more than two two-level atoms the subradiant Dicke states are merely non-symmetric \cite{Dicke(1954)}; the corresponding states have been recently used to form a unimodular basis \cite{Vetter(2016), Jen(2016)}. Various theoretical investigations have discussed the preparation \cite{Maser(2009),Ammon(2012),Bienaime(2012),Genes(2015),Tang(2015),Scully(2015),Mirza(2016),Durand(2016),Damanet(2016),Bettles(2016), Ganesh(2016), Jen(2017),Hebenstreit(2017),Guerin(2017)} as well as the subradiant emission characteristics of non-symmetric Dicke states for larger atomic ensembles, either using a semiclassical theory \cite{Bienaime(2012), Durand(2016),Damanet(2016), Bettles(2016), Ganesh(2016), Jen(2017), Hebenstreit(2017),Guerin(2017)} or within a full quantum mechanical treatment \cite{Tang(2015),Scully(2015),Mirza(2016),Wiegner(2011)PRA}. 
Very recently, the first experimental observation of 
retarded subradiant spontaneous decay for more than two emitters has been reported \cite{Guerin(2016)}.

Most theoretical studies of subradiant systems have investigated the temporal aspects of subradiance,  
only few have have been devoted to its particular spatial emission properties 
\cite{Wiegner(2011)PRA,Tang(2015), Durand(2016)}. Yet, in correspondence to their superradiant counterparts,
distant sources prepared in fully antisymmetric or non-symmetric Dicke states display pronounced directional emission profiles, e.g., exhibiting a strong suppression of spontaneous radiation in particular directions \cite{Wiegner(2011)PRA}. 


In this letter, we study the spatial emission characteristics of 
light sources arranged in totally antisymmetric states, thereby investigating what we call \textit{directional Dicke subradiance}. 
We start to analyze the conditions for achieving totally antisymmetric Dicke states for $N \geq 2$  multi-level atoms or multi-photon sources and    
explore the specific spatial emission profiles of emitters prepared in such states. 
We next discuss the possibilities to observe subradiant directional emission behavior of classical sources.
In particular, we show that the same multi-photon interferences and thus the same subradiant suppression of incoherent radiation derived for quantum emitters can be obtained with thermal light sources (TLS),
if projected into particular correlated states via photon detection. 
Finally, we present measurements of directional Dicke subradiance for up to five TLS.



In general, a totally antisymmetric state of $N$ sources is defined by 
\begin{equation}
\ket{A_{N}} = \frac{1}{\sqrt{N!}} \sum_{\mathcal{P}} \text{sgn}(\mathcal{P}) \ket{n_{\mathcal{P}_1},n_{\mathcal{P}_2},\ldots,n_{\mathcal{P}_N}} \ ,
\label{eq:AntiState}
\end{equation}
where $\ket{n_{l}}$ describes the state of source $l$, $l = 1, \ldots, N$, and $\sum_\mathcal{P}$ represents the sum over all permutations of the sources, with $\text{sgn}(\mathcal{P})$ the sign of the permutation.
According to Eq.~(\ref{eq:AntiState}), in order to realize a totally antisymmetric state, all sources have to be prepared in distinct states, i.e., $n_{i}\neq n_{j}$, for $i\neq j$, otherwise the state vanishes.
In particular, in the case of two-level atoms, 
a fully antisymmetric state only exists for $N=2$ particles.
Thus constructing totally antisymmetric states for $N>2$ emitters requires internal level schemes with at least $N$ distinguishable states, e.g., an excited state $\ket{e}$ and $N-1$ distinguishable  ground states $\ket{g_{\tilde{l}}}$, $\tilde{l}=1,\hdots,N-1$ \cite{Hebenstreit(2017)}.
We call this kind of source \textit{multi-level single photon emitter} (MSPE).
To construct a totally antisymmetric state for $N$ MSPE we 
choose 
$n_{1}=e,\, n_{2}=g_{1}, n_{3}=g_{2}, \ldots,\, n_{N}=g_{N-1}$ .

\begin{figure}[b]
\centering
\includegraphics[width=0.7\columnwidth]{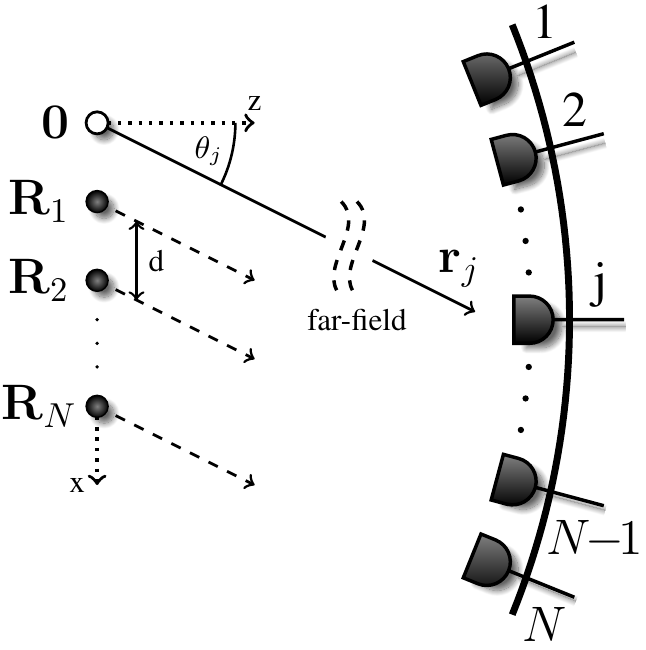}
\caption{Setup considered for the oberservation of directional Dicke subradiance: $N$ light sources are aligned along the x-axis equidistantly at positions $\vec{R}_{l}$, $l=1,\hdots,N$, with separation $d\gg \lambda$. In the far field of the sources, $N$ detectors measure the $N$th-order correlation function at positions $\vec{r}_{j}$, $j=1,\hdots,N$. 
}
\label{fig:setup}
\end{figure}

To determine the spatial emission characteristics of such a state we assume without loss of generality the simple source arrangement shown in Fig.~\ref{fig:setup}, where $N$ MSPE are located equidistantly with separation $d$ along the x-axis at positions $\vec{R}_{l} = l \, d \, \vec{e}_x$,  $l=1,\ldots , N$.
The intensity recorded at position $\vec{r}_{1}$ is defined by  
\begin{equation}
	I(\vec{r}_{1}) = G^{(1)}_{\rho}(\vec{r}_{1}) = \left< E^{(-)}(\vec{r}_{1}) E^{(+)}(\vec{r}_{1}) \right>_{\rho} \ ,
\label{eq:DefIntensity}
\end{equation}
where $\rho$ is the density matrix of the $N$ MSPE and the (dimensionless) positive frequency part of the electric field operator in the far field of the sources is given by \cite{Wiegner(2015)}
\begin{equation}
\begin{aligned}
	\left[ E^{(-)}(\vec{r}_1) \right]^\dagger = E^{(+)}(\vec{r}_1) \propto \sum_{l=1}^{N} e^{-i\,l\, \delta_{1}} \;\hat{S}_{-}^{(l)} \; .
\end{aligned}
\label{eq:DefE1}
\end{equation}
In Eq.~(\ref{eq:DefE1}), $\delta_1= 
- k \, d \,  \sin(\theta_{1})$ corresponds to the relative phase of a photon emitted by source $l$ and recorded by a detector at $\vec{r}_{1}$ with respect to a photon emitted at the origin (see Fig.~\ref{fig:setup}), and $\hat{S}_{-}^{(l)}=\frac{1}{\sqrt{N-1}}\sum_{\tilde{l}=1}^{N-1}\hat{s}_{-}^{(l,\tilde{l})}$ is the sum over all atomic lowering operators $\hat{s}_{-}^{(l,\tilde{l})} = \ket{g_{\tilde{l}}}_{l}\!\bra{e}$, $\tilde{l} = 1, \ldots, N-1$, deexciting the $l$th MSPE from its upper state $\ket{e}_{l}$ to the ground state $\ket{g_{\tilde{l}}}_{l}$. 

For $N$ MSPE in the antisymmetric state $\ket{A_{N}}$ with one excitation, i.e., with $\sum_{l=1}^{N}\braket{ \hat{S}^{(l)}_{+} \hat{S}^{(l)}_{-}}_{\ket{A_{N}}}= 1$, the intensity as a function of $\delta_{1}$  calculates 
to
\begin{align}
\label{eq:intensity}
	&I_{\ket{A_{N}}}\left(\delta_{1}\right) = \bra{A_{N}} E^{(-)}(\vec{r}_{1}) E^{(+)}(\vec{r}_{1}) \ket{A_{N}} \nonumber \\[1mm]
	&=  \sum_{l=1}^{N} \left< \hat{S}^{(l)}_{+} \hat{S}^{(l)}_{-} \right>_{\ket{A_{N}}} + \sum_{\substack{ l_{1}, l_{2} = 1 \\ l_{1} \neq l_{2}}}^{N} e^{i\, \delta_{1} (l_{1}-l_{2})} \left< \hat{S}^{(l_{1})}_{+} \hat{S}^{(l_{2})}_{-} \right>_{\ket{A_{N}}}  \nonumber \\
	&= \frac{N}{N-1} \left[ 1 - \frac{1}{N^2} \frac{\sin\left(\frac{N\delta_{1}}{2}\right)^{2}}{\sin\left(\frac{\delta_{1}}{2}\right)^{2}} \right]  \, ,
\end{align}
where in Eq.~(\ref{eq:intensity}) we exploited the fact that all interference terms contribute with equal weight, i.e., $\braket{\hat{S}^{(l_{1})}_{+} \hat{S}^{(l_{2})}_{-}}_{\ket{A_{N}}}=-1/[N(N-1)]$, for $l_{1} \neq l_{2}$. Note that this equality is also expressed in the cross-correlation coefficient \cite{Agarwal(2014)}, which is identical for all source pairs $l_{1} \neq l_{2}$
\begin{equation}
\label{eq:correlationsAnti}
\frac{\braket{\hat{S}^{(l_{1})}_{+} \hat{S}^{(l_{2})}_{-}}_{\ket{A_{N}}}}{\sqrt{\braket{\hat{S}^{(l_{1})}_{+} \hat{S}^{(l_{1})}_{-}}_{\ket{A_{N}}}\braket{\hat{S}^{(l_{2})}_{+} \hat{S}^{(l_{2})}_{-}}_{\ket{A_{N}}}}}=-\frac{1}{N-1} \ .
\end{equation}

Eq.~(\ref{eq:intensity}) displays an intensity profile marked by pronounced dips of vanishing radiation, analogous to the inverted intensity profile  of a coherently illuminated  grating with $N$ slits. Note that pronounced peaks in the intensity characterized by a grating function are  well known for 
$N$ two-level atoms in \textit{symmetric} Dicke states, arranged in the same manner as in Fig.~\ref{fig:setup} \cite{Wiegner(2011)PRA}. Such \textit{superradiant} emission profiles can further be observed when recording higher-order intensity correlation functions for both,  $N$ uncorrelated fully excited two-level atoms \cite{Oppel(2014),Wiegner(2015)} and $N$ uncorrelated classical light sources \cite{Oppel(2014),Bhatti(2016)}. Yet, in contrast to the sharp peaks of \textit{increased} intensity  in the case of superradiant emission, the inverted grating function of Eq.~(\ref{eq:intensity}) reveals highly focused dips of \textit{reduced} intensity, i.e., directional Dicke subradiance. Equally to its superradiant counterpart 
\cite{Wiegner(2011)PRA,Oppel(2014),Wiegner(2015),Bhatti(2016)}, the subradiant intensity profile of Eq.~(\ref{eq:intensity}) for $N$ MSPE in the antisymmetric state $\ket{A_{N}}$ with one excitation displays a visibility of  $\mathcal{V}=1$, with the minima located at $\delta_{1} = 2 m \pi$, $m \in \mathbb{Z}$, having an angular width of $\delta\theta_{1} \approx 2\pi / (N k d)$.
Note that these properties describe distinctive features of superradiance \cite{Wiegner(2011)PRA,Oppel(2014),Wiegner(2015),Bhatti(2016)} but can also be used, as in our case, to characterize the particular 
attributes of Dicke subradiance.

A further option to construct totally antisymmetric states $\ket{A_N}$ and observe the corresponding subradiant behavior is to make use of multi-photon sources (MPS). Hereby, each source $l$ emits a discrete number of photons $n_{l}$, 
assumed to be different from the other sources, i.e., $n_{i}\neq n_{j}$, for $i\neq j$.
This could be realized, e.g., by combining $n_{l}$ single photon emitters for each source $l$.
Again considering the source arrangement of Fig.~\ref{fig:setup}, the (dimensionless) positive frequency part of the electric field operator in the far field of the sources 
reads \cite{Bhatti(2016)}
\begin{equation}
\begin{aligned}
	\left[ E^{(-)}(\vec{r}_1) \right]^\dagger = E^{(+)}(\vec{r}_1) \propto \sum_{l=1}^{N} e^{-i\,l\, \delta_{1}} \;\hat{a}_{l} ,
\label{eq:DefE2}
\end{aligned}
\end{equation}
where $\hat{a}_{l}$ denotes the annihilation operator of a photon emitted from the $l$th MPS.
Choosing $n_{1}=0, n_{2}=1, \ldots, n_{N}=N-1$ 
yields again the intensity profile of Eq.~(\ref{eq:intensity}), i.e., a distribution following a negative grating function and displaying directional Dicke subradiance, yet with a different global prefactor of $N^{2}/2$ 
\cite{AdditionalInformation}.

Note that when computing the $N$th-order intensity correlation function $G^{(N)}_{N\, TLS}(\delta_{1},\hdots, \delta_{N})$ of a light field produced by $N$ TLS,
similar multi-photon interference terms appear as those occuring in the derivation of $I_{\ket{A_{N}}}\left(\delta_{1}\right)$ for $N$ MPS \cite{AdditionalInformation}. This indicates how directional subradiant behavior can be observed also with classical light sources, i.e., by exploiting correlations produced among TLS when recording a specific number of photons at particular positions \cite{Oppel(2014),Bhatti(2016)}.



To corroborate this argument we consider again the source arrangement of Fig.~\ref{fig:setup}, where this time  
$N$ detectors are placed at positions $\vec{r}_j$, $j=1, \ldots , N$, in the far field of $N$ TLS. The density matrix $\rho_{N\, TLS}$ of the field  generated by the $N$ TLS can be written in the number-state representation in the form \cite{Oppel(2014),Bhatti(2016)}
\begin{equation}
	\rho_{N\, TLS} = \bigotimes_{l=1}^{N} \sum_{n_{l}=0}^{\infty} P_{TLS}(n_{l}) \ket{n_{l}}\bra{n_{l}} \ ,
\label{eq:rho_TLS}
\end{equation}
where $P_{TLS}(n_{l})$ denotes the (Bose-Einstein) distribution of source $l$, and
we assume equal mean photon numbers for all sources, i.e., $\bar{n} = \bar{n}_{l} = \braket{\hat{a}_{l}^{\dagger}\hat{a}_{l}}_{\rho}$, $l=1,\hdots, N$.

The $N$th-order intensity correlation function for $N$ light sources is defined by \cite{Glauber(1963)Quantum}
\begin{equation}
\begin{aligned}
	G^{(N)}_{N}(\vec{r}_{1},\hdots, \vec{r}_{N}) = \left<: \prod_{j=1}^{N} E^{(-)}(\vec{r}_{j}) E^{(+)}(\vec{r}_{j}) : \right>_{\rho_{N}} ,
\label{eq:DefGN}
\end{aligned}
\end{equation}
where $\left< : F :\right>_{\rho_{N}}$ represents the (normally ordered) quantum mechanical expectation value of the operator $F$ for a field in the state $\rho_{N}$.


To observe directional Dicke subradiance 
via measurements of $G^{(N)}_{N}(\vec{r}_{1},\hdots, \vec{r}_{N})$ we suppose that $N-1$ detectors are placed at the fixed \textit{subradiant positions} (SP)
\begin{equation}
\delta_{j}=2 \pi \frac{(j-1)}{N} \ \ , \ \  j=2,\hdots ,N \ .
\label{eq:AntiPos}
\end{equation}
These positions are identical to the arguments of the $N$th roots of unity and therefore fulfill the identity
\begin{equation}
\label{eq:mthroots}
	\sum_{j=2}^{N}e^{i\delta_{j}n}=\begin{cases} \ \ -1 \ \ &, n\neq\{0\} \ , \ \text{mod}(N)\\(N\!-\!1)  \ &, n=\{0\} \ , \ \text{mod}(N) \, . \end{cases}
\end{equation}
Since we assume $N$ statistically independent TLS, we can make use of the Gaussian moment theorem to write Eq.~(\ref{eq:DefGN}) also in the form
\begin{equation}
\label{eq:GNTLS}
	G^{(N)}_{N\, TLS}(\delta_{1},\hdots, \delta_{N}) = \sum_{\mathcal{P}}\prod_{j=1}^{N} \left< E^{(-)}(\delta_{j}) E^{(+)}(\delta_{\mathcal{P}_{j}}) \right>  ,
\end{equation}
where now $\sum_{\mathcal{P}}$ refers to the sum over all permutations 
of the $N$ detectors, and
the first moment is given by [cf. Eq.~(\ref{eq:DefE2})]
\begin{equation}
\begin{aligned}
	\left< E^{(-)}(\delta_{j_1}) E^{(+)}(\delta_{j_2}) \right> =  \bar{n} \sum_{l=1}^{N} e^{i l (\delta_{j_{1}}-\delta_{j_{2}})} \ ,
\label{eq:crosscorr}
\end{aligned}
\end{equation}
where the label $\rho_{N\,TLS}$ of the expectation value has been dropped for simplicity.

In the case that detector $\delta_{1}$ is not involved in the sum, i.e., for $j_{1/2}=2,\hdots , N$, Eq.~(\ref{eq:crosscorr}) simplifies to [cf. Eq.~(\ref{eq:mthroots})]
\begin{align}
\label{eq:crosscorr2}
	& \left< E^{(-)}(\delta_{j_1}) E^{(+)}(\delta_{j_2}) \right>  =  \bar{n} \left[ 1 + \sum_{l=2}^{N} e^{i \delta_{l} (j_1 - j_{2})} \right] \nonumber \\
	 & =\begin{cases} \ \ 0 \ \ &, (j_1 - j_{2}) \neq\{0\} \ , \ \text{mod}(N) \\ N  \bar{n}  \ &, (j_1 - j_{2}) =\{0\} \ , \ \text{mod}(N) \, , \end{cases}
\end{align}
which means that all cross-correlation terms of any two fixed detectors $j_1 \neq j_{2}$ vanish. 
Eq.~(\ref{eq:GNTLS}) thus reduces to 
\begin{equation}
\begin{aligned}
	&G^{(N)}_{N\, TLS}(\delta_{1},\text{SP}) =  \prod_{j=1}^{N}  \left< E^{(-)}(\delta_{j}) E^{(+)}(\delta_{j}) \right> \\
		&+ \sum_{k =2}^{N} \left| \left< E^{(-)}(\delta_{1}) E^{(+)}(\delta_{k}) \right> \right|^{2}	\prod_{\stackrel{j=2}{j\neq k }}^{N}\left< E^{(-)}(\delta_{j}) E^{(+)}(\delta_{j}) \right>  \, ,
\label{eq:GNTLS22}
\end{aligned}
\end{equation}
where the interference term 
is given by [cf. Eq.~(\ref{eq:mthroots})]
\begin{align}
\label{eq:interference}
	& \sum_{k =2}^{N} \left| \left< E^{(-)}(\delta_{1}) E^{(+)}(\delta_{k}) \right> \right|^{2} = \sum_{k =2}^{N} \left|  \bar{n} \sum_{l=1}^{N} e^{i l (\delta_{1}-\delta_{k})} \right|^{2} \nonumber \\ 
	& = \bar{n}^{2}\, (N-1) \sum_{l_{1},l_{2}=1}^{N} e^{i\delta_{1}(l_{1}-l_{2})} - \bar{n}^{2} \sum_{ \substack{l_{1},l_{2} = 1\\ l_{1}\neq l_{2}}}^{N} e^{i\delta_{1}(l_{1}-l_{2})} \nonumber \\
	& = \bar{n}^{2} N^{2} \left[ 1 - \frac{1}{N^2} \frac{\sin\left(\frac{N\delta_{1}}{2}\right)^{2}}{\sin\left(\frac{\delta_{1}}{2}\right)^{2}} \right] \, .
\end{align}
The normalized $N$th-order intensity correlation function
finally reads
\begin{equation}
\begin{aligned}
	g^{(N)}_{N\, TLS} (\delta_{1},\text{SP})  = 2 - \frac{1}{N^{2}} \frac{\sin\left(\frac{N\delta_{1}}{2}\right)^{2}}{\sin\left(\frac{\delta_{1}}{2}\right)^{2}} \ ,
\label{eq:gNfinal}
\end{aligned}
\end{equation}
displaying an inverted grating function identical to the one obtained for quantum sources in Eq.~(\ref{eq:intensity}), i.e., a subradiant intensity distribution $I_{\ket{A_{N}}}\left(\delta_{1}\right)$ produced by MSPE and MPS in the totally antisymmetric state $\ket{A_{N}}$.
Note that due to the different constant term the visibility of the classical subradiant pattern 
of Eq.~(\ref{eq:gNfinal}) equals $\mathcal{V}_{TLS}=1/3$, independently of the number of sources $N$.

To reach higher visibilities for classical sources one could increase the number of fixed detectors, e.g., using multiples $\alpha$ of complete sets of $N-1$ detectors placed at the SP, i.e., $\alpha(N-1)$, $\alpha \in \mathbb{N}_{+}$. In this case we obtain 
\cite{AdditionalInformation}
\begin{equation}
\begin{aligned}
	 g^{(1+\alpha(N-1))}_{N\, TLS} (\delta_{1},\alpha\! \times\! \text{SP})  \sim  1 + \alpha - \frac{\alpha}{N^{2}} \frac{\sin\left(\frac{N\delta_{1}}{2}\right)^{2}}{\sin\left(\frac{\delta_{1}}{2}\right)^{2}}  ,
\label{eq:gNafinal}
\end{aligned}
\end{equation}
where in the limit $\alpha\gg 1$ the visibility $\mathcal{V}_{TLS}^{(\alpha)}=\alpha/(\alpha+2)$ approaches unity, as in the case of quantum sources.


\begin{figure}[b]
	\centering
		\includegraphics[width=1.0\columnwidth]{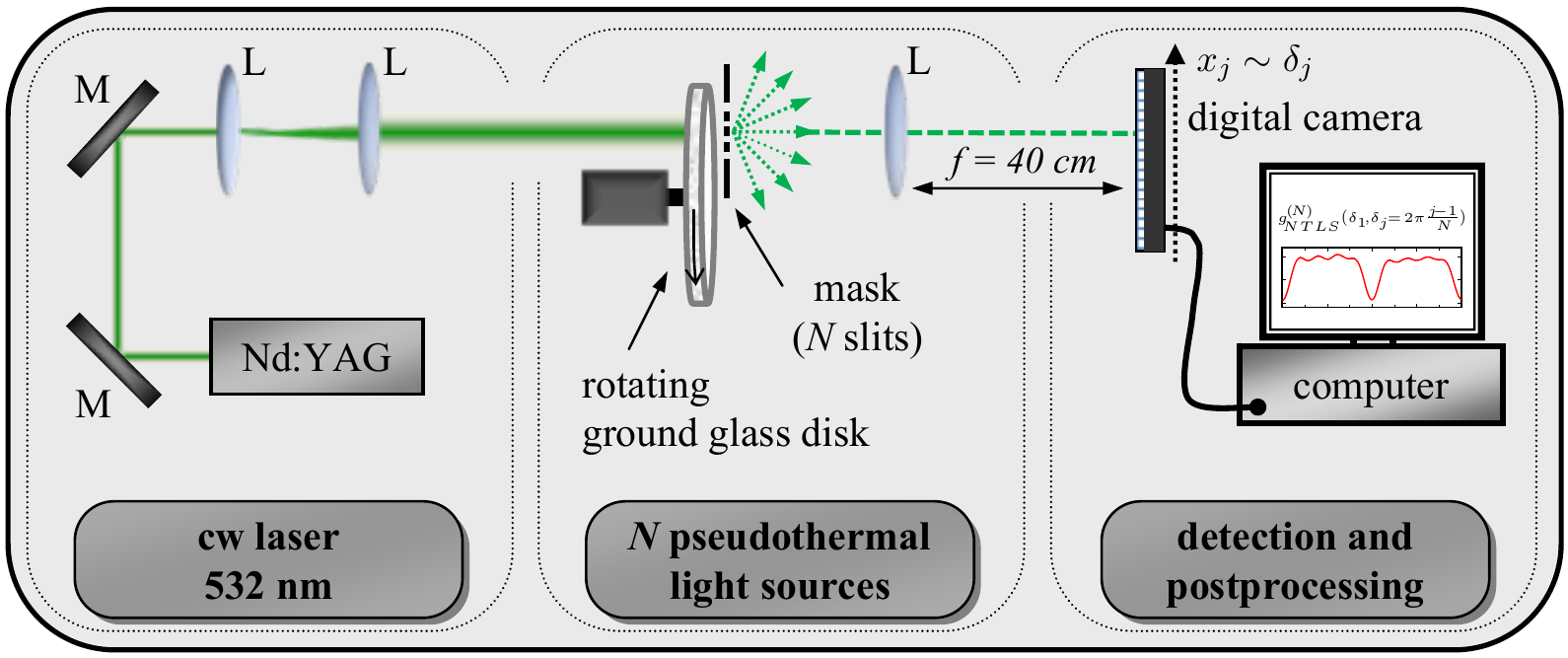}
	\caption{
(Color online) Experimental  setup  to  measure $g^{(N)}_{N\, TLS}$ with $N$ pseudothermal light sources. M: mirror, L: lens.	For details see text and Ref.~\cite{Oppel(2014)}.}
	\label{fig:Experiment}
\end{figure}


Note that recently 
an isomorphism between $G^{(m)}_{\rho_{N}}$ and $G^{(1)}_{\tilde{\rho}_{N}^{(m-1)}}$, $m \in \mathbb{N}_{+}$, was identified for a light field $\rho_{N}$ produced by $N$ sources \cite{Wiegner(2015),Bhatti(2016)}, i.e.,
\begin{equation}
	G^{(m)}_{\rho_{N}}(\delta_1, \hdots,\delta_m) = G_{\tilde{\rho}_{N}^{(m-1)}}^{(1)}(\delta_{1})\, G_{\rho_{N}}^{(m-1)}(\delta_{2},\hdots,\delta_{m})  ,
\label{eq:Gm_rho_G1}
\end{equation}
where $\tilde{\rho}_{N}^{(m-1)}$ describes the state of the field after $m-1$ photons have been recorded at positions $\delta_{2},\hdots, \delta_{m}$. In the case of $N$ TLS and $\alpha(N-1)$ detectors placed at the SP the projected state $\tilde{\rho}_{N\, TLS}^{(\alpha(N-1))}$ reads
\begin{equation}
\label{eq:thermal_state_matrice}
\tilde{\rho}_{N\, TLS}^{(\alpha(N-1))}\!= \frac{ \left[\prod_{j=2}^{N} E^{(+)}(\delta_{j}) \right]^{\alpha}\!\! \rho_{N\, TLS}   \left[\prod_{j=2}^{N} E^{(-)}(\delta_{j}) \right]^{\alpha} }{G_{\rho_{N\, TLS}}^{(\alpha(N-1))}(\alpha\!\times\!\text{SP})} ,
\end{equation}
with $\text{Tr}[\tilde{\rho}_{N\, TLS}^{(\alpha(N-1))}] = 1$. The state $\tilde{\rho}_{N\, TLS}^{(\alpha(N-1))}$ is not of diagonal form, where the nondiagonal terms describe the correlations between the TLS induced by the detection of $\alpha(N-1)$ photons at the SP.
The corresponding cross-correlation coefficient 
is given by \cite{AdditionalInformation}
\begin{equation}
\label{eq:correlationsTLS}
	\frac{\left< \hat{a}^{\dagger}_{l_{1}} \hat{a}_{l_{2}} \right>_{\tilde{\rho}_{N\, TLS}^{(\alpha(N-1))}}}{\sqrt{\left< \hat{a}^{\dagger}_{l_{1}} \hat{a}_{l_{1}} \right>_{\tilde{\rho}_{N\, TLS}^{(\alpha(N-1))}}\left< \hat{a}^{\dagger}_{l_{2}} \hat{a}_{l_{2}} \right>_{\tilde{\rho}_{N\, TLS}^{(\alpha(N-1))}}}} = - \frac{\alpha}{N+N\alpha-\alpha} \, ,
\end{equation}
demonstrating that the cross correlations 
are identical for any two sources $l_{1}\neq l_{2}$.
In particular, 
for 
$\alpha \gg N$, Eq.~(\ref{eq:correlationsTLS}) becomes identical to Eq.~(\ref{eq:correlationsAnti}), displaying the correlations between any two of $N$ multi-level atoms prepared in the totally antisymmetric state ${\ket{A_{N}}}$.

\begin{figure}[t]
	\centering
		\includegraphics[width=1.0\columnwidth]{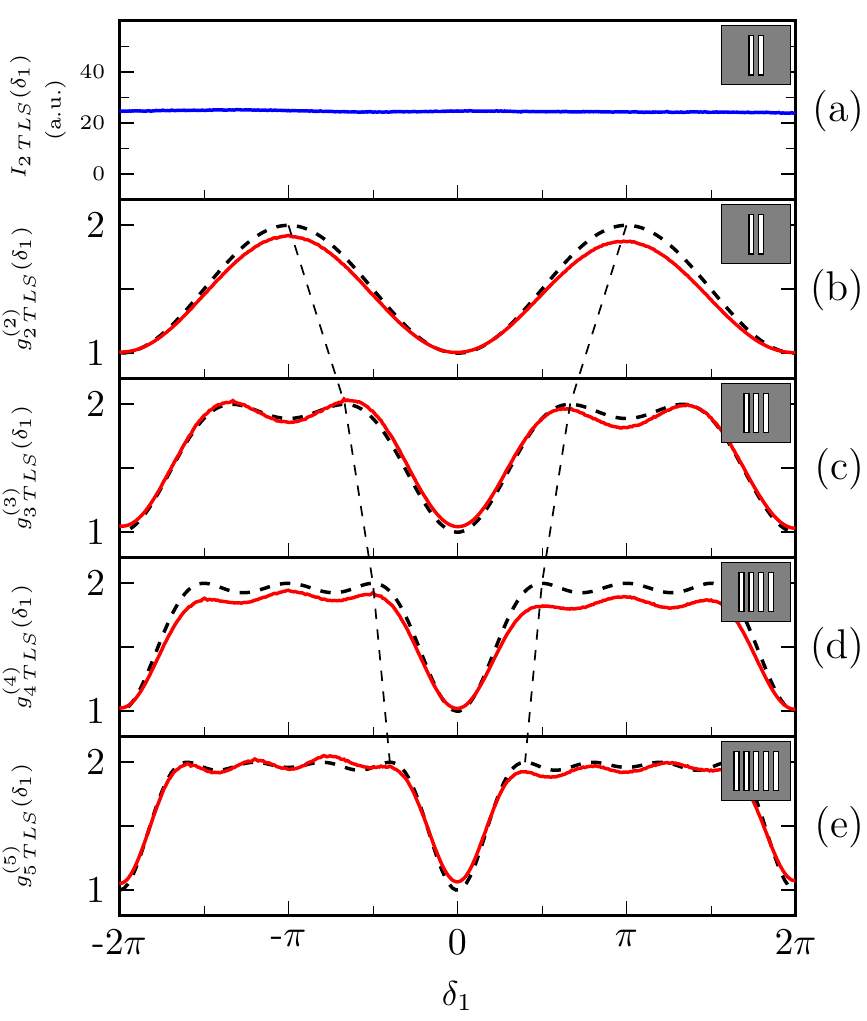}
	\caption{(Color online) Experimental results. (a) Average intensity $I_{2 \, TLS} (\delta_{1})$ of $N=2$ TLS demonstrating that the pseudothermal light sources are spatially incoherent in first order. (b)-(e) Measurement of the normalized $N$th order correlation function $g^{(N)}_{N\, TLS} (\delta_{1},\delta_{j}\!=\!2\pi\frac{j-1}{N}; j\!=\!2,\hdots ,N) = \mathrm{G}^{(N)}_{N \, TLS}(\delta_{1},\delta_{j}\!=\!2\pi\frac{j-1}{N}; j\!=\!2,\hdots ,N)/[I_{N \, TLS} (\delta_{1}) \prod_{j = 2}^{N} I_{N \, TLS} (\delta_{j}\!=\!2\pi\frac{j-1}{N})]$ for $N = 2, \ldots, 5$ as a function of the first detector at $\delta_{1}$ (red solid curves). The suppression of incoherently emitted radiation at $\delta_{1} = 0$ after $N - 1$ photons have been recorded at the SP at $\delta_{j}\!=\!2\pi\frac{j-1}{N}$, $j=2,\hdots ,N$, is clearly visible. The theoretical predictions of Eq.~(\ref{eq:gNfinal}) are displayed by the black (dashed) curves.
	}
	\label{fig:Measurement}
\end{figure}

To measure $g^{(N)}_{N\, TLS}$ for $N$ statistically independent TLS we use the pseudothermal light of a coherently illuminated rotating ground glass disk \cite{Estes(71)} (coherence time $\tau_{c}\approx50$ms), impinging on a mask with $N$ identical slits of width $a=25\mu$m and separation $d=200\mu$m (see Fig.~\ref{fig:Experiment}). As coherent light source we utilize a linearly polarized frequency-doubled Nd:YAG laser at $\lambda=532$nm. Working in the high intensity regime we employ a conventional digital camera placed in the far field of the mask to determine $g^{(N)}_{N\, TLS}(\delta_1, \hdots,\delta_N)$,
where we correlate $N-1$ pixels of the camera located at the SP $\sim \delta_{2},\hdots, \delta_{N}$ with one moving pixel $\sim \delta_{1}$ (integration time of the camera
$\tau_{i}\approx 1 \text{ms} \ll \tau_{c}$) \cite{Oppel(2014)}.

The experimental results
for $g^{(N)}_{N\, TLS}(\delta_1, \hdots,\delta_N)$ obtained in this way for $N=2,\hdots,5$ TLS are shown in Fig.~\ref{fig:Measurement}. From the plots the directional Dicke subradiant behavior of the $N$ TLS is clearly visible, with dips of vanishing radiation that are
in excellent agreement in position, depth and width with the theoretical predictions of Eq.~(\ref{eq:gNfinal}). This confirms the theory outlined above stating that a) with distant quantum sources as well as classical TLS in corresponding states
directional subradiance can be observed  [cf. Eqs.~(\ref{eq:intensity}) and (\ref{eq:gNafinal})], and that b) directional subradiance of TLS occurs
due to the similarity between the totally antisymmetric state ${\ket{A_{N}}}$ of $N$ quantum sources and the highly correlated state $\tilde{\rho}_{N\, TLS}^{(\alpha(N-1))}$, obtained from $N$ initially uncorrelated TLS via $\alpha(N-1)$ photon detection events at the SP [cf. Eqs.~(\ref{eq:AntiState}) and (\ref{eq:thermal_state_matrice})], leading to identical cross correlations for $N$ quantum sources and for $N$ TLS in the limit $\alpha\gg N$ [cf. Eqs.~(\ref{eq:correlationsAnti}) and (\ref{eq:correlationsTLS})].

In conclusion we presented a detailed discussion of the spatial aspects of Dicke subradiance, i.e., the intensity profiles observed for the incoherently emitting distant sources prepared in corresponding states.
We examined the conditions to achieve totally antisymmetric states for  multi-level atoms or multi-photon sources and derived analytical expressions for the resulting spatial spontaneous emission patterns. We also showed that directional Dicke subradiance can be observed with incoherently emitting TLS, i.e., by measuring higher order photon correlations projecting the TLS into highly correlated states. The latter is an unexpected outcome as subradiance is considered to be a purely
nonlocal, nonclassical phenomenon displayed by quantum sources \cite{Hebenstreit(2017)}.

%
%

\begin{acknowledgments}
The authors thank G. S. Agarwal and S. M\"ahrlein for helpful comments and fruitfull discussions.
The authors gratefully acknowledge funding by the Erlangen Graduate School in Advanced Optical Technologies (SAOT) by the German Research Foundation (DFG) in the framework of the German excellence initiative. D.B. gratefully acknowledges financial support by the Cusanuswerk, Bisch\"ofliche Studienf\"orderung.
\end{acknowledgments}


%

\newpage

\begin{widetext}

\section{Supplemental Material: Directional Dicke Subradiance with Nonclassical and Classical Sources}

\renewcommand\theequation{S\arabic{equation}}

\section{Antisymmetric Quantum State: Multi-level Single Photon Emitters}

We investigate totally antisymmetric quantum states, defined by [see Eq.~(1) in the main text]
\begin{align}
\ket{A_{N}} = \frac{1}{\sqrt{N!}} \sum_{\mathcal{P}} \text{sgn}(\mathcal{P}) \ket{n_{\mathcal{P}_1},n_{\mathcal{P}_2},\ldots,n_{\mathcal{P}_N}} .
\label{eq:antistate}
\end{align}
To fulfill the requirement $n_{i}\neq n_{j}$, $i \neq j$,
we introduced  in the main text multi-level single photon emitters (MSPE) with one excited state $\ket{e}$ and $N-1$ ground states $\ket{g_{\tilde{l}}}$, $\tilde{l}=1,\ldots,N-1$, where we choose $n_{1}=e, n_{2}=g_{1}, \ldots, n_{N}=g_{N-1}$.

For the $l$th atom the transition from the excited state to one of the ground states is given by a superposition of lowering operators $\hat{s}_{-}^{(l,\tilde{l})}=\ket{g_{\tilde{l}}}_{l}\bra{e}$. The total lowering operator $\hat{S}_{-}^{(l)}$ of the $l$th atom thus reads
\begin{align}
\hat{S}_{-}^{(l)}=  \frac{1}{\sqrt{N-1}}\left( \ket{g_{1}}_{l} + \ket{g_{2}}_{l} + \cdots + \ket{g_{N-1}}_{l} \right) \otimes_{\ l \! \!} \bra{e} =\frac{1}{\sqrt{N-1}}\sum_{\tilde{l}=1}^{N-1}\hat{s}_{-}^{(l,\tilde{l})} \ .
\label{eq:LoweringOp}
\end{align}

Using Eq.~(\ref{eq:LoweringOp}) and the definition of the (dimensionless) positive frequency part of the electric field operator in the far field for the sources [see Eq.~(3) in main text]
\begin{align}
	\left[ E^{(-)}(\vec{r}_1) \right]^\dagger = E^{(+)}(\vec{r}_1) \propto \sum_{l=1}^{N} e^{-i\,l\, \delta_{1}} \;\hat{S}_{-}^{(l)} \; ,
\label{eq:DefE12}
\end{align}
we can  calculate the intensity $	I_{\ket{A_{N}}}\left(\delta_{1}\right)$ of $N$ MSPE with a single excitation prepared in the arrangement of Fig. 1 of the main text in the antisymmetric state $\ket{A_{N}}$ 
\begin{align}
\label{eq:IntensityMSPE}
	I_{\ket{A_{N}}}\left(\delta_{1}\right) = \bra{A_{N}} E^{(-)}(\vec{r}_{1}) E^{(+)}(\vec{r}_{1}) \ket{A_{N}} &=  \sum_{l=1}^{N} \left< \hat{S}^{(l)}_{+} \hat{S}^{(l)}_{-} \right>_{\ket{A_{N}}} + \sum_{\substack{ l_{1}, l_{2} = 1 \\ l_{1} \neq l_{2}}}^{N} e^{i\, \delta_{1} (l_{1}-l_{2})} \left< \hat{S}^{(l_{1})}_{+} \hat{S}^{(l_{2})}_{-} \right>_{\ket{A_{N}}}  \nonumber \\[1mm]
	&= \frac{N}{N-1} \left[ 1 - \frac{1}{N^{2}} \frac{\sin\left(\frac{N\delta_{1}}{2}\right)^{2}}{\sin\left(\frac{\delta_{1}}{2}\right)^{2}} \right] ,
\end{align}
where in Eq.~(\ref{eq:IntensityMSPE}) we exploited the fact that if the state $\ket{A_{N}}$ contains a single excitation we have
\begin{align}
\label{eq:correlationsAnti1}
\sum_{l=1}^{N}\left< \hat{S}^{(l)}_{+} \hat{S}^{(l)}_{-}\right>_{\ket{A_{N}}}= N \left< \hat{S}^{(1)}_{+} \hat{S}^{(1)}_{-}\right>_{\ket{A_{N}}} = \frac{N}{N!} \sum_{\mathcal{P}} \bra{e,n_{\mathcal{P}_{2}},\ldots,n_{\mathcal{P}_{N}}} \frac{1}{N-1} \sum_{\tilde{l}=1}^{N-1}\left( \hat{s}_{+}^{(1,\tilde{l})}\hat{s}_{-}^{(1,\tilde{l})} \right) \ket{e,n_{\mathcal{P}_{2}},\ldots,n_{\mathcal{P}_{N}}} = 1  ,
\end{align}
and that in Eq.~(\ref{eq:IntensityMSPE}) all $N-1$ interference terms of two arbitrary sources $l_{1} \neq l_{2}$ contribute with equal weight
\begin{align}
\label{eq:correlationsAnti2}
\left< \hat{S}^{(l_{1})}_{+} \hat{S}^{(l_{2})}_{-}\right>_{\ket{A_{N}}} &= \left< \hat{S}^{(1)}_{+} \hat{S}^{(2)}_{-}\right>_{\ket{A_{N}}} \nonumber \\
&= - \frac{1}{N!} \sum_{\tilde{l}=1}^{N-1} \sum_{\mathcal{P}} \bra{e,g_{\tilde{l}},n_{\mathcal{P}_{3}},\ldots,n_{\mathcal{P}_{N}}} \frac{\hat{s}_{+}^{(1,\tilde{l})}\hat{s}_{-}^{(2,\tilde{l})}}{N-1} \ket{g_{\tilde{l}},e,n_{\mathcal{P}_{3}},\ldots,n_{\mathcal{P}_{N}}} = -\frac{1}{N(N-1)} \,  .
\end{align}
Note that in Eqs.~(\ref{eq:correlationsAnti1}) and (\ref{eq:correlationsAnti2}) we took into account that the state $\ket{A_{N}}$  sums over all possible permutations [see  Eq.~(\ref{eq:antistate})].

One can now calculate the cross correlation coefficient [see main text Eq.~(5)]
\begin{align}
\frac{\left<\hat{S}^{(l_{1})}_{+} \hat{S}^{(l_{2})}_{-}\right>_{\ket{A_{N}}}}{\sqrt{\left<\hat{S}^{(l_{1})}_{+} \hat{S}^{(l_{1})}_{-}\right>_{\ket{A_{N}}}\left<\hat{S}^{(l_{2})}_{+} \hat{S}^{(l_{2})}_{-}\right>_{\ket{A_{N}}}}}=-\frac{1}{N-1}\, , \  \text{with} \ l_{1}\neq l_{2} \, ,
\label{eq:CrossCorrCoeffMSPE}
\end{align}
where we have $\left< \hat{S}^{(l_{1})}_{+} \right>_{\ket{A_{N}}} = \left< \hat{S}^{(l_{2})}_{-} \right>_{\ket{A_{N}}} = 0$.

\section{Antisymmetric Quantum State: Multiphoton Sources}

A second type of quantum source discussed in the main text are multiphoton sources (MPS), able to emit any discrete number of photons $n$. To generate antisymmetric states $\ket{A_{N}}$ from $N$ MPS the emitted photon numbers are chosen to be $n_{1}=0, n_{2}=1, \ldots, n_{N}=N-1$, fulfilling the requirement $n_{i} \neq n_{j}$ for all $i \neq j$.

Using the electric field operator at position $\vec{r}_{1}$ in the far field of the sources
\begin{align}
	\left[ E^{(-)}(\vec{r}_1) \right]^\dagger = E^{(+)}(\vec{r}_1) \propto \sum_{l=1}^{N} e^{-i\,l\, \delta_{1}} \;\hat{a}_{l} ,
\label{eq:DefE22}
\end{align}
with the bosonic lowering operator $\hat{a} \ket{n} = \sqrt{n} \ket{n-1}$, the intensity in the far field of $N$ MPS prepared in the arrangement of Fig. 1 of the main text calculates to
\begin{align}
\label{eq:IntensityMPS}
	I_{N\, MPS}\left(\delta_{1}\right) = \bra{A_{N}} E^{(-)}(\vec{r}_{1}) E^{(+)}(\vec{r}_{1}) \ket{A_{N}} &=  \sum_{l=1}^{N} \left< \hat{a}^{\dagger}_{l} \hat{a}_{l} \right>_{\ket{A_{N}}} + \sum_{\substack{ l_{1}, l_{2} = 1 \\ l_{1} \neq l_{2}}}^{N} e^{i\, \delta_{1} (l_{1}-l_{2})} \left< \hat{a}^{\dagger}_{l_{1}} \hat{a}_{l_{2}} \right>_{\ket{A_{N}}}  \nonumber \\[1mm]
	&= \frac{N^{2}}{2} \left[ 1 - \frac{1}{N^{2}} \frac{\sin\left(\frac{N\delta_{1}}{2}\right)^{2}}{\sin\left(\frac{\delta_{1}}{2}\right)^{2}} \right] \, ,
\end{align}
where in Eq.~(\ref{eq:IntensityMPS}) the following moments have been used:
\begin{align}
\sum_{l=1}^{N} \left< \hat{a}^{\dagger}_{l} \hat{a}_{l} \right>_{\ket{A_{N}}} = N \left< \hat{a}^{\dagger}_{1} \hat{a}_{1} \right>_{\ket{A_{N}}} = \frac{N}{N!} \sum_{\tilde{l}=1}^{N} \sum_{\mathcal{P}} \bra{n_{\tilde{l}},n_{\mathcal{P}_{2}},\ldots ,n_{\mathcal{P}_{N}}} \hat{a}^{\dagger}_{1} \hat{a}_{1} \ket{n_{\tilde{l}},n_{\mathcal{P}_{2}},\ldots ,n_{\mathcal{P}_{N}}} = \sum_{n=0}^{N-1} n = \frac{N(N-1)}{2} \, ,
\label{moment1}
\end{align}
and
\begin{align}
\left< \hat{a}^{\dagger}_{l_{1}} \hat{a}_{l_{2}} \right>_{\ket{A_{N}}} = \left< \hat{a}^{\dagger}_{1} \hat{a}_{2} \right>_{\ket{A_{N}}} &= - \frac{1}{N!} \sum_{\tilde{l}=1}^{N-1} \sum_{\mathcal{P}} \bra{n_{\tilde{l}+1},n_{\tilde{l}},n_{\mathcal{P}_{3}},\ldots ,n_{\mathcal{P}_{N}}} \hat{a}^{\dagger}_{1} \hat{a}_{2} \ket{n_{\tilde{l}},n_{\tilde{l}+1},n_{\mathcal{P}_{3}},\ldots ,n_{\mathcal{P}_{N}}} \nonumber \\[1mm]
& = - \frac{(N-2)!}{N!} \sum_{n=1}^{N-1}n = - \frac{1}{2} \, .
\label{moment2}
\end{align}
Note that Eq.~(\ref{eq:IntensityMPS}) has a minimal value of zero and thus displays a visibility $\mathcal{V}=1$.

Using Eqs.~(\ref{moment1}) and (\ref{moment2}) the cross correlation coefficient can be computed:
\begin{align}
\frac{\left< \hat{a}_{l_{1}}^{\dagger} \hat{a}_{l_{2}} \right>_{\ket{A_{N}}}}{\sqrt{\left< \hat{a}_{l_{1}}^{\dagger} \hat{a}_{l_{1}} \right>_{\ket{A_{N}}} \left< \hat{a}_{l_{2}}^{\dagger} \hat{a}_{l_{2}} \right>_{\ket{A_{N}}}}} = - \frac{1}{N-1} \, , \ \text{with} \ l_{1}\neq l_{2}\, ,
\end{align}
where again we have $\big< \hat{a}_{l_{1}}^{\dagger} \big>_{\ket{A_{N}}} = \big< \hat{a}_{l_{2}} \big>_{\ket{A_{N}}} = 0$.
Note that this expression is identical to the cross correlation coefficient  calculated for $N$ MSPE [cf. Eq.~(\ref{eq:CrossCorrCoeffMSPE})].

Finally, it is also possible to calculate the normalized maximum of the intensity distribution, when integrating the intensity given in Eq.~(\ref{eq:IntensityMPS}) over one period of the moving detector phase $\delta_{1}$
\begin{align}
		\left< I_{N\, MPS} \right>_{\delta_{1}} = \frac{1}{2\pi} \int  I_{N\, MPS}(\delta_{1})\, d\delta_{1} = \frac{N(N-1)}{2} \ .
\end{align}
This leads to the following normalized maximum of the intensity distribution
\begin{align}
\left. I_{N\, MPS}/\left< I_{N\, MPS} \right>_{\delta_{1}} \right|_{max} = \frac{N}{N-1} \, ,
\end{align}
which is identical to the maximum of the intensity distribution of $N$ MSPE prepared in an antisymmetric state  $\ket{A_{N}}$ [see Eq.~(\ref{eq:IntensityMSPE})]. 

\section{Thermal light sources}

Finally, we want to calculate the $(1+\alpha(N-1))$th order intensity correlation function $G_{N\, TLS}^{(1+\alpha(N-1))}(\delta_{1},\alpha \! \times \! \text{SP})$ for $N$ thermal light sources (TLS) in the arrangement of Fig.~1 (see main text), with $\alpha(N-1)$ detectors placed at the subradiant positions (SP) 
\begin{align}
\delta_{j}=2\pi \frac{ j-1}{N}\, , \ \text{with} \ j=2,\ldots,N \, ,
\end{align}
i.e., with $\alpha$ detectors at each of the SP, and a single moving detector at $\delta_{1}$, to create directional subradiance with a visibility approaching unity
(see main text and the result in Eq.~(17)).
The $(1+\alpha(N-1))$th order correlation function with one moving detector and $\alpha$ detectors at each of the SP is defined by
\begin{align}
	G_{N\, TLS}^{(1+\alpha(N-1))}(\delta_{1},\alpha \! \times \! \text{SP}) = \sum_{\mathcal{P}}  \prod_{j=1}^{1+\alpha(N-1)} \left< E^{(-)}(\delta_{j}) E^{(+)}(\delta_{\mathcal{P}_{j}}) \right> \, ,
	\label{eq:gaussian_moment_1}
\end{align}
where the Gaussian moment theorem has been employed.
Due to the mathematical properties of the SP [cf. Eq.~(13) in the main text] and the definition of the electric field operator [cf. Eq.~(\ref{eq:DefE22})], only such field cross correlations survive in Eq.~(\ref{eq:gaussian_moment_1}), which include $\delta_{1}$ or which correlate fields at the same detector position:
\begin{align}
	&G^{(1+\alpha(N-1))}_{N\, TLS}(\delta_{1},\alpha \! \times \! \text{SP}) =  \left< E^{(-)}(\delta_{1}) E^{(+)}(\delta_{1}) \right> (\alpha !)^{N-1} \left( \prod_{j=2}^{N}  \left< E^{(-)}(\delta_{j}) E^{(+)}(\delta_{j}) \right> \right)^{\alpha} \nonumber \\
	&+ \alpha^{2} \sum_{k =2}^{N} \left| \left< E^{(-)}(\delta_{1}) E^{(+)}(\delta_{k}) \right> \right|^{2} (\alpha -1)! \, (\alpha !)^{N-2}	\left( \prod_{\stackrel{j=2}{j\neq k }}^{N}\left< E^{(-)}(\delta_{j}) E^{(+)}(\delta_{j}) \right> \right) \left( \prod_{j=2}^{N}\left< E^{(-)}(\delta_{j}) E^{(+)}(\delta_{j}) \right> \right)^{\alpha -1} \nonumber\\
	&= (\alpha !)^{N-1} \left< E^{(-)}(\delta_{j}) E^{(+)}(\delta_{j}) \right>^{\alpha(N-1)-1} \left[ \left< E^{(-)}(\delta_{j}) E^{(+)}(\delta_{j}) \right>^{2} + \alpha \, \bar{n}^{2} \left[ N^{2} - \frac{\sin\left(\frac{N\delta_{1}}{2}\right)^{2}}{\sin\left(\frac{\delta_{1}}{2}\right)^{2}} \right] \right]  \, ,
\label{eq:G1+a(N-1)}
\end{align}
where in line 3 of Eq.~(\ref{eq:G1+a(N-1)}) we made use of the result of Eq.~(15) in the main text.
$G^{(1+\alpha(N-1))}$ can be normalized by use of $G^{(1)}_{N\, TLS}(\delta_{j}) = N \bar{n}$, which leads to
\begin{align}
	&g^{(1+\alpha(N-1))}_{N\, TLS}(\delta_{1},\alpha \! \times \! \text{SP}) = (\alpha !)^{N-1} \left[ 1 + \alpha  - \frac{\alpha}{N^{2}} \frac{\sin\left(\frac{N\delta_{1}}{2}\right)^{2}}{\sin\left(\frac{\delta_{1}}{2}\right)^{2}} \right]   \, ,
\end{align}
where the maximum and the minimum are given by $g^{(1+\alpha(N-1))}_{N\, TLS \, max}(\delta_{1},\alpha \! \times \! \text{SP}) = (1+\alpha) (\alpha !)^{N-1}$ and $g^{(1+\alpha(N-1))}_{N\, TLS \, min}(\delta_{1},\alpha \! \times \! \text{SP}) = (\alpha !)^{N-1}$, respectively, so that the visibility calculates to $\mathcal{V}_{TLS}^{(\alpha)}= \frac{\alpha}{\alpha+2}$, i.e., approaching unity $\mathcal{V}_{TLS}^{(\alpha \gg 1)} \rightarrow 1$ for  $\alpha \gg 1$. 
Clearly it can be seen that for $\alpha =1$ one obtains the already known result of $g_{N\, TLS}^{(N)}$ (cf. Eq.~(16) in the main text), with a maximum $g_{N\, TLS\, max}^{(N)} = 2$, a  minimum $ g_{N\, TLS\, min}^{(N)} =1$, and a visibility $\mathcal{V}_{TLS}^{(1)}=1/3$, independently of the number of sources $N$. Note that it is rather unexpected that for $\alpha >1$ the visibility remains independent of the number of sources, i.e., $\mathcal{V}_{TLS}^{(\alpha)}$ is solely determined by $\alpha$.

Integrating over one period of the moving detector phase $\delta_{1}$
\begin{align}
	\left< g^{(1+\alpha(N-1))}_{N\, TLS} \right>_{\delta_{1}} = \frac{1}{2\pi} \int  g^{(1+\alpha(N-1))}_{N\, TLS}(\delta_{1},\alpha \! \times \! \text{SP})\, d\delta_{1} = (\alpha !)^{N-1} \frac{N+N\alpha-\alpha}{N} \, ,
\end{align}
one can compute the normalized maximum
\begin{align}
\left. g^{(1+\alpha(N-1))}_{N\, TLS}/\left< g^{(1+\alpha(N-1))}_{N\, TLS} \right>_{\delta_{1}} \right|_{max} = \frac{N+N\alpha}{N+N\alpha-\alpha},
\end{align}
and the normalized minimum
\begin{align}
\left. g^{(1+\alpha(N-1))}_{N\, TLS}/\left< g^{(1+\alpha(N-1))}_{N\, TLS} \right>_{\delta_{1}} \right|_{min} = \frac{N}{N+N\alpha-\alpha}.
\end{align}
For $\alpha\gg N$ the maximum scales as $\sim N/(N-1)$ while the minimum converges towards $\sim 0$. This is identical to the outcome obtained for the quantum case, i.e., the maximum and the minimum of the intensity $	I_{\ket{A_{N}}}\left(\delta_{1}\right)$ of $N$ MSPE with a single excitation [cf. Eq.~(\ref{eq:IntensityMSPE})].

The cross-correlation coefficient in the case of TLS is given by
\begin{align}
C^{(N,\alpha)}=\frac{\left< \hat{a}_{l_{1}}^{\dagger} \hat{a}_{l_{2}} \right>_{\tilde{\rho}_{N}^{(\alpha(N-1))}}}{\sqrt{\left< \hat{a}_{l_{1}}^{\dagger} \hat{a}_{l_{1}} \right>_{\tilde{\rho}_{N}^{(\alpha(N-1))}} \left< \hat{a}_{l_{2}}^{\dagger} \hat{a}_{l_{2}} \right>_{\tilde{\rho}_{N}^{(\alpha(N-1))}}}} \, ,  \ \text{with} \ l_{1}\neq l_{2}\, ,
\label{eq:CNa}
\end{align}
with $\tilde{\rho}_{N}^{(\alpha(N-1))}$ denoting the state of the $N$ TLS after $\alpha(N-1)$ photons have been detected at the SP:
\begin{align}
\label{eq:thermal_state_matrice2}
\tilde{\rho}_{N}^{(\alpha(N-1))}= \frac{ \left[\prod_{j=2}^{N} E^{(+)}(\delta_{j}) \right]^{\alpha} \rho_{N}  \left[\prod_{j=2}^{N} E^{(-)}(\delta_{j}) \right]^{\alpha} }{G_{N\, TLS}^{(\alpha(N-1))}(\alpha \!\times\!\text{SP})} \, ,
\end{align}
where again we have $\big< \hat{a}_{l_{1}}^{\dagger} \big>_{\tilde{\rho}_{N}^{(\alpha(N-1))}} = \big< \hat{a}_{l_{2}} \big>_{\tilde{\rho}_{N}^{(\alpha(N-1))}} = 0$.

Employing again the Gaussian moment theorem, the cross-correlation term in Eq.~(\ref{eq:CNa}) calculates to:
\begin{align}
	\left< \hat{a}_{l_{1}}^{\dagger} \hat{a}_{l_{2}} \right>_{\tilde{\rho}_{N}^{(\alpha(N-1))}} &  =\frac{1}{G_{N\, TLS}^{(\alpha(N-1))}(\alpha \!\times\!\text{SP})}\, \alpha^{2} \sum_{k=2}^{N}\left< \hat{a}_{l_{1}}^{\dagger}\, E^{(+)}(\delta_{k}) \right> \left< E^{(-)}(\delta_{k})\, \hat{a}_{l_{2}} \right> \nonumber \\[1mm]
	&\phantom{=} \times (\alpha-1)! (\alpha !)^{N-2} \left( \prod_{\stackrel{j=2}{j\neq k }}^{N}\left< E^{(-)}(\delta_{j}) E^{(+)}(\delta_{j}) \right> \right) \left( \prod_{j=2}^{N}\left< E^{(-)}(\delta_{j}) E^{(+)}(\delta_{j}) \right> \right)^{\alpha-1} \nonumber \\[0mm]
	&=\frac{1}{G_{N\, TLS}^{(\alpha(N-1))}(\alpha \!\times\!\text{SP})}\, \alpha (\alpha !)^{N-1} (N \bar{n})^{\alpha(N-1)-1} \bar{n}^{2} \sum_{k=2}^{N} e^{i l_{1} \delta_{k}} e^{- i l_{2} \delta_{k}} \nonumber \\[2mm]
	&= -\frac{\alpha \, (\alpha !)^{N-1} \, N^{\alpha(N-1)-1} \, \bar{n}^{\alpha(N-1)+1}}{(\alpha !)^{N-1} (N\bar{n})^{\alpha(N-1)}}  \, ,
\end{align}
\begin{align}
\rightarrow \left< \hat{a}_{l_{1}}^{\dagger} \hat{a}_{l_{2}} \right>_{\tilde{\rho}_{N}^{(\alpha(N-1))}} = - \alpha \frac{\bar{n}}{N} \, , \ \text{with} \ l_{1}\neq l_{2} \, .
\label{eq:al1al2TLS}
\end{align}
To derive Eq.~(\ref{eq:al1al2TLS}) the $\alpha(N-1)$th-order correlation function $G_{N\, TLS}^{(\alpha(N-1))}(\alpha \!\times\!\text{SP})$ has been introduced, leaving out the moving detector at $\delta_{1}$. $G_{N\, TLS}^{(\alpha(N-1))}(\alpha \!\times\!\text{SP})$ is identical to the normalization of the $\alpha(N-1)$-photon subtracted state of $N$ TLS given in Eq.~(\ref{eq:thermal_state_matrice2}), with $\alpha$ times $(N-1)$ detectors placed at the SP, and can be computed to
\begin{align}
	G^{(\alpha(N-1))}_{N\, TLS}(\alpha \! \times\! \text{SP})= (\alpha !)^{N-1} \left[ \prod_{j=2}^{N} \left< E^{(-)}(\delta{j})E^{(+)}(\delta{j}) \right> \right]^{\alpha} = (\alpha !)^{N-1} (N\bar{n})^{\alpha(N-1)}\, .
\end{align}

On the other hand, the normalizing intensities in Eq.~(\ref{eq:CNa}) are given by:
\begin{align}
	& \left< \hat{a}_{l_{1}}^{\dagger} \hat{a}_{l_{1}} \right>_{\tilde{\rho}_{N}^{(\alpha(N-1))}}  = \frac{1}{G_{N\, TLS}^{(\alpha(N-1))}(\alpha \!\times\!\text{SP})}\left[ \left< \hat{a}_{l_{1}}^{\dagger} \hat{a}_{l_{1}} \right> (\alpha !)^{N-1} \left( \prod_{j=2}^{N}\left< E^{(-)}(\delta_{j}) E^{(+)}(\delta_{j}) \right> \right)^{\alpha} \right.\nonumber \\[2mm]
	&\left. + \alpha^{2} \sum_{k=2}^{N}\left< \hat{a}_{l_{1}}^{\dagger}\, E^{(+)}(\delta_{k}) \right> \left< E^{(-)}(\delta_{k})\, \hat{a}_{l_{1}} \right>  (\alpha-1)! (\alpha !)^{N-2} \left( \prod_{\stackrel{j=2}{j\neq k }}^{N}\left< E^{(-)}(\delta_{j}) E^{(+)}(\delta_{j}) \right> \right) \left( \prod_{j=2}^{N}\left< E^{(-)}(\delta_{j}) E^{(+)}(\delta_{j}) \right> \right)^{\alpha-1} \right] \nonumber \\[1mm]
	&=\frac{1}{G_{N\, TLS}^{(\alpha(N-1))}(\alpha \!\times\!\text{SP})}\left[ (\alpha !)^{N-1} N^{\alpha(N-1)} \bar{n}^{\alpha(N-1)+1} + \alpha (\alpha !)^{N-1} (N \bar{n})^{\alpha(N-1)-1} \bar{n}^{2} \sum_{k=2}^{N} 1 \right] \nonumber \\[2mm]
	&=\frac{ (\alpha !)^{N-1} \,  N^{\alpha(N-1)} \, \bar{n}^{\alpha(N-1)+1}}{(\alpha !)^{N-1} (N\bar{n})^{\alpha(N-1)}}  \left[ 1 + \alpha - \frac{\alpha}{N}  \right]  ,
\end{align}
\begin{align}
\rightarrow \left< \hat{a}_{l_{1}}^{\dagger} \hat{a}_{l_{1}} \right>_{\tilde{\rho}_{N}^{(\alpha(N-1))}} = \left< \hat{a}_{l_{2}}^{\dagger} \hat{a}_{l_{2}} \right>_{\tilde{\rho}_{N}^{(\alpha(N-1))}} = \bar{n} \left[ 1 + \alpha - \frac{\alpha}{N} \right]  .
\end{align}
This leads to the following simple expression for the cross-correlation coefficient [cf. Eq~(\ref{eq:CNa})]
\begin{align}
	C^{(N,\alpha)} = - \frac{\alpha}{N[1+\alpha -\alpha/N]} = - \frac{\alpha}{N + N\alpha -\alpha} .
\label{eq:CrossCorrCoeffTLS}
\end{align}
In particular, in the limit $\alpha \gg N$, we find that Eq.~(\ref{eq:CrossCorrCoeffTLS}) scales as $\sim - 1/(N-1)$, which is identical to the cross correlation coefficient of $N$ MSPE prepared in the antisymmetric state $\ket{A_{N}}$ [cf. Eq.~(\ref{eq:CrossCorrCoeffMSPE})].

\end{widetext}

\end{document}